\documentclass[pra,aps,twocolumn,showpacs,floatfix]{revtex4-1}
\usepackage{amssymb}
\usepackage{graphicx}
\usepackage{dcolumn}
\usepackage{bm,amsmath,verbatim}
\usepackage{mathrsfs}
\usepackage{color}
\usepackage[colorlinks=true, letterpaper=true, pdfstartview=FitV, linkcolor=red, citecolor=blue, urlcolor=red]{hyperref}

\def\be{\begin{equation}}
\def\ee{\end{equation}}
\def\bea{\begin{eqnarray}}
\def\eea{\end{eqnarray}}
\def\bse{\begin{subequations}}
\def\ese{\end{subequations}}

\def\be{\begin{eqnarray}}
\def\ee{\end{eqnarray}}

\begin{document}
\title{Bright solitons in a 2D spin-orbit-coupled dipolar Bose-Einstein condensate}
\author{Yong Xu$^{1}$}
\author{Yongping Zhang$^{2}$}
\author{Chuanwei Zhang$^{1}$}\email{chuanwei.zhang@utdallas.edu}
\affiliation{$^{1}$Department of Physics, The University of Texas at Dallas, Richardson,
Texas 75080, USA}
\affiliation{$^{2}$Quantum Systems Unit, OIST Graduate University, Onna, Okinawa 904-0495, Japan}
\begin{abstract}
We study a two-dimensional spin-orbit-coupled dipolar Bose-Einstein condensate with repulsive
contact interactions by both the variational method and the imaginary time evolution of the
Gross-Pitaevskii equation. The dipoles are completely polarized along one direction
in the 2D plane so as to provide an effective attractive dipole-dipole interaction.
We find two types of solitons as the ground states arising from such attractive interactions:
a plane wave soliton with a spatially varying phase and a stripe soliton with a spatially
oscillating density for each component. Both types of solitons possess smaller
size and higher anisotropy than the soliton without spin-orbit coupling. Finally, we discuss
the properties of moving solitons, which are nontrivial because of the violation of Galilean invariance.
\end{abstract}
\maketitle

\section{introduction}
Ever since the first achievement of Bose-Einstein condensates (BECs) in ultracold atomic gases~\cite{BookSmith},
matter wave solitons have been the central focus of many experimentalists and theorists~\cite{KevrekidisBook}.
Solitons are the result of the interplay between nonlinearity and dispersion and keep their shape while
traveling. In BECs, nonlinearity originates from collisional interactions between
atoms, which can be readily tuned via Feshbach resonances~\cite{Chin2010RMP}.
In general, there are two types of solitons: a bright soliton with a density bump for attractive interactions
and a dark soliton with a density notch and a phase jump for repulsive interactions. Both bright and dark solitons
have been experimentally observed in cold atoms with contact interactions~\cite{Burger1999PRL,Phillips2000Science,Cornell2001PRL,Hulet2002Nature,Salomon2002Science,%
Wieman2006PRL,Oberthaler2004PRL,Peter2013PRL,Hulet2014Nature}.
However, for such contact attractive interactions, bright solitons can only exist in one dimension (1D),
but not in two dimensions (2D) where the states either collapse or expand~\cite{note0}.

Different from the local nonlinearity resulting from contact interactions, the non-local nonlinearity
can stabilize a 2D bright soliton~\cite{Wyller2001PRE,Pfau2009RPP}, in particular, the nonlinearity
introduced by the dipole-dipole interaction.
This interaction is long ranged and anisotropic with the strength
and sign (i.e. repulsive or attractive) depending on the dipole orientation.
When an external rotating magnetic field is applied to reverse the sign of the interaction~\cite{Santos2005PRL},
or the dipoles are completely polarized in a 2D plane~\cite{Malomed2008PRL}, the dipolar interaction
can become attractive and 2D bright solitons can be, therefore, stabilized under appropriate conditions.
It is essential to note that although the relevant interaction in common experiments with cold atomic
gases is contact, increasing interest has been focused on the atoms with large magnetic moments
that possess dipole-dipole interactions~\cite{Pfau2009RPP,Bohn2011PRL,Zoller2012}.
In fact, the Bose-Einstein condensation of several dipolar atoms
such as Chromium~\cite{Pfau2005PRL,Pfau2008NatPhys,Beaufils2008PRA},
Dysprosium~\cite{Lev2011PRL}, and Erbium~\cite{Ferlaino2012PRL},
as well as the degeneracy of a dipolar Fermi gas~\cite{Lev2012PRL,Ferlaino2014PRL} have been observed
in experiments.

Recently, the spin-orbit coupling between two hyperfine states of cold atoms has been experimentally engineered
~\cite{Lin2011Nature,Jing2012PRL,Zwierlen2012PRL,PanJian2012PRL,Peter2013,Spilman2013PRL}.
And this achievement has ignited tremendous interest in this field because of the dramatic change
in the single particle dispersion (induced by spin-orbit coupling) which in conjunction with the interaction
leads to many exotic superfluids~\cite{Galitski2008PRA,ZhaiHui2010PRL,Wu2011CPL,Ho2011PRL,Santos2011PRL,Yongping2012PRL,%
Hu2012PRL,Li2012PRL,Chen2012PRA,Kuei2015arXiv,Yong2015arXiv}(also see~\cite{Ohberg2011RMP,Spilman2013NatRev,Xiangfa2013JPB,%
Goldman2014RPP,Zhai2015RPP,WeiArxiv,Jing2014,Yong2015JP} for review).
Such change in dispersion also results in exotic solitons even when the interaction is contact,
including bright solitons~\cite{Santos2010PRL,Yong2013PRA,Achilleos2014PRL,Salasnich,Malomed2014PRE,Malomed2014PRA,Sakaguchi},
dark solitons~\cite{Fialko,Achilleos2013EL}, and gap solitons~\cite{Kartashov2013PRL,Kartashov2014PRL,Yongping2015arXiv}
for BECs, as well as dark solitons for Fermi superfluids~\cite{Yong2014Soliton,XJ2015PRA}. These solitons exhibit unique
features that are absent without spin-orbit coupling, for instance, the plane wave profile with a spatially varying
phase and the stripe profile with a spatially oscillating density for BECs, as well as
the presence of Majorana fermions inside a soliton for Fermi superfluids.
Also, the violation of Galilean invariance~\cite{Qizhong2013,Yong2013PRA,QizhongReview} by spin-orbit coupling dictates that the
structure of solitons changes with their velocities.

On the other hand, spin-orbit-coupled BECs with dipole-dipole
interactions~\cite{Yi2012PRL,Cui2013PRA,Ng2014PRA,Yousefi2014arXiv} have also been explored,
and intriguing quasicrystals~\cite{Demler2013PRL} as well as meron states~\cite{Clark2013PRL} have been found. However, whether a soliton
can exist in such BECs in 2D with long ranged dipole-dipole interactions and spin-orbit dispersion has not yet been investigated.

In this paper, we examine the existence and properties of a bright soliton in a
two species spin-orbit-coupled dipolar BEC in 2D with repulsive contact interactions via both
the variational method and the imaginary time evolution of the Gross-Pitaevskii equation (GPE).
The dipoles are completely oriented along the $y$ direction in the 2D plane in
order to provide an effective attractive dipole-dipole interaction. Thanks to such
attractive interactions, we find two types of solitons: a plane wave soliton
(when the repulsive intraspecies contact interaction is larger than the repulsive interspecies one)
and a stripe soliton (when the interspecies one is larger). These 2D solitons
as the ground states cannot exist for a system with pure attractive contact interactions and spin-orbit coupling.
Such solitons are highly anisotropic and their size is also
reduced by spin-orbit coupling. Finally, we study the moving solitons, which are nontrivial because of the
lack of Galilean invariance. The size of a soliton first increases and then decreases with the rise of the velocity
and this change is anisotropic. The moving soliton also tends to be plane wave even when its
stationary counterpart has the stripe structure.

The paper is organized as follows. In Sec. II, we introduce the energy functional and
the time-dependent GPE, which are used to describe a spin-orbit-coupled dipolar BEC. In Sec. III,
we calculate the bright soliton by performing the minimization of the energy of the variational
ansatz wave functions and an imaginary time evolution of the GPE. The properties of such soliton are also
explored by the former method. Then, we study the nontrivial moving solitons in Sec. IV. Finally,
we conclude in Sec. V.

\section{Model}
We consider a Rashba-type spin-orbit-coupled BEC and write its single particle
Hamiltonian as
\begin{equation}
H_s=\frac{\hat{\bf p}^2}{2m}+\frac{1}{2}m\omega_\perp^2\rho^2+\frac{1}{2}m\omega_z^2z^2+\lambda(\hat{{\bf p}}\times{\bm \sigma})\cdot {\bf e}_z,
\end{equation}
where $\hat{\bf p}=-i\hbar\nabla$ is the momentum operator, $m$ is the atom mass, $\lambda$ is the spin-orbit coupling strength,
and $\bm \sigma$ are Pauli matrices. $\omega_\perp$ ($\omega_z$) is the
trap frequency in the $(x,y)$ plane (along the $z$ direction). Here, we assume that
$\hbar\omega_z$ is much larger than $\hbar\omega_\perp$ and the mean-field interaction so that
the atoms are frozen to the ground state in the $z$ direction. Given that a soliton
is studied, we thus set $\omega_\perp=0$.

When the $s$-wave contact and dipole-dipole interactions are involved, the energy functional of a
2D condensate can be written as
\begin{eqnarray}
E=&&\int d {\bf r} \left[\Psi({\bf r})^\dagger H_s\Psi({\bf r})+\frac{1}{2}g(|\Psi_\uparrow|^4+
|\Psi_\downarrow|^4)\right. \nonumber \\
&&\left. +g_{12}|\Psi_\uparrow|^2 |\Psi_\downarrow|^2 \right]+E_{dd},
\label{energyFunction}
\end{eqnarray}
where the condensate wave function $\Psi({\bf r})=[\Psi_\uparrow({\bf r}),\Psi_\downarrow({\bf r})]^T$
with two pseudo-spin components $\Psi_{\uparrow (\downarrow)}({\bf r})$
, $g=4\pi\hbar^2 a/(\sqrt{2\pi}l_z m)$ and $g_{12}=4\pi\hbar^2 a_{12}/(\sqrt{2\pi}l_z m)$
are the intraspecies and interspecies contact interaction strength respectively with the intraspecies and
interspecies $s$-wave scattering length being $a$ and $a_{12}$ and the characteristic length along
$z$ being $l_z=\sqrt{\hbar/(m\omega_z)}$. Here, $H_s=-\hbar^2(\partial_x^2+\partial_y^2)/(2m)-i\hbar\lambda(\partial_x\sigma_y-\partial_y\sigma_x)$
is the 2D single particle Hamiltonian, and the third dimension has been integrated out.
For dipole-dipole interactions, we only consider the density-density
interaction which is dominant when a two subspace (i.e. two pseudo-spin states) of a large spin atom (e.g. dysprosium) is considered.
We also assume that the dipoles are all oriented along the $y$ direction, thus
\begin{equation}
E_{dd}=\frac{g_d}{2}\frac{1}{(2\pi)^2}\int d{\bf k} \rho_{\bf k}\rho_{-{\bf k}}U_{d}({\bf k}l_z),
\end{equation}
where the Fourier transform of the total density is $\rho_{\bf k}=\int d{\bf r} e^{-i{\bf k}\cdot{\bf r}}
(|\Psi_\uparrow|^2+|\Psi_\downarrow|^2)$ and $U({\bf k})$ is given by
\begin{equation}
U_d({\bf k}l_z)=-\sqrt{2\pi}+\frac{3\pi l_z k^2_y e^{k^2 l_z^2/2}\text{erfc}(kl_z/\sqrt{2})}{k},
\end{equation}
with $\text{erfc}$ being the complementary error function. Here, $g_d=\mu_0 \mu^2/(6\pi l_z)$ characterizes
the strength of the dipole-dipole interaction where $\mu$ is the magnetic dipolar moment and $\mu_0$ is
the permeability of the free space.

The dynamical behavior of a BEC can be described by the time-dependent GPE
\begin{equation}
i\hbar\frac{\partial\Psi({\bf r})}{\partial t}=H_s\Psi({\bf r})+g G\Psi({\bf r})
+g_d U_d({\bf r})\Psi({\bf r}),
\end{equation}
where the contact interaction matrix is
\begin{eqnarray}
G=\left(
  \begin{array}{cc}
    |\Psi_\uparrow|^2+\frac{g_{12}}{g}|\Psi_\downarrow|^2 & 0 \\
    0 & |\Psi_\downarrow|^2+\frac{g_{12}}{g}|\Psi_\uparrow|^2 \\
  \end{array}
\right),
\end{eqnarray}
and the dipolar interaction potential is
\begin{equation}
U_d({\bf r})=\frac{1}{(2\pi)^2}\int d{\bf k} e^{i{\bf k}\cdot{\bf r}}\rho({\bf k})U_d({\bf k}l_z).
\end{equation}

For numerical simulation, we choose $\hbar\omega_z$, $l_z$, and $1/\omega_z$ as the units of
energy, length, and time, respectively, and the dimensionless energy per atom hence reads
\begin{eqnarray}
\mathcal{E}=&&\int d {\bf r} \left[\Phi({\bf r})^\dagger \mathcal{H}_s\Phi({\bf r})+\frac{1}{2}\gamma(|\Phi_\uparrow|^4+
|\Phi_\downarrow|^4)\right. \nonumber \\
&&\left. +\gamma_{12}|\Phi_\uparrow|^2 |\Psi_\downarrow|^2 \right]+\frac{\gamma_d}{2(2\pi)^2}\int d{\bf k} n_{\bf k}n_{-{\bf k}}U_{d}({\bf k}),
\label{energy_dim}
\end{eqnarray}
where $\mathcal{H}_s=-(\partial_x^2+\partial_y^2)/2-i\alpha(\partial_x\sigma_y-\partial_y\sigma_x)$,
$\alpha=\lambda/(\omega_x l_z)$,
$\gamma=2\sqrt{2\pi}N_0 a/l_z$, $\gamma_{12}=2\sqrt{2\pi}N_0 a_{12}/l_z$ with the total particle number $N_0$,
, $\gamma_d=2N_0 a_d/l_z$ with $a_{d}=m\mu_0\mu^2/(12\pi\hbar^2)$, and
$n_{\bf k}=\int d{\bf r} e^{-i{\bf k}\cdot{\bf r}} (|\Phi_\uparrow|^2+|\Phi_\downarrow|^2)$.
The wave function is normalized to 1 (i.e. $\int d{\bf r}(|\Phi_\uparrow|^2+|\Phi_\downarrow|^2)=1$).

The dimensionless time-dependent GPE reads
\begin{eqnarray}
i\frac{\partial\Phi({\bf r})}{\partial t}=&&{\mathcal H}_s\Phi({\bf r})+\gamma \mathcal{G}\Phi({\bf r}) \nonumber \\
&&+\frac{\gamma_d}{(2\pi)^2}\int d{\bf k} e^{i{\bf k}\cdot{\bf r}}n({\bf k})U_d({\bf k}) \Phi({\bf r}),
\label{GP_time}
\end{eqnarray}
where
\begin{eqnarray}
\mathcal{G}=\left(
  \begin{array}{cc}
    |\Phi_\uparrow|^2+\frac{\gamma_{12}}{\gamma}|\Phi_\downarrow|^2 & 0 \\
    0 & |\Phi_\downarrow|^2+\frac{\gamma_{12}}{\gamma}|\Phi_\uparrow|^2 \\
  \end{array}
\right).
\end{eqnarray}

\begin{figure*}[t]
\includegraphics[width=5.2in ]{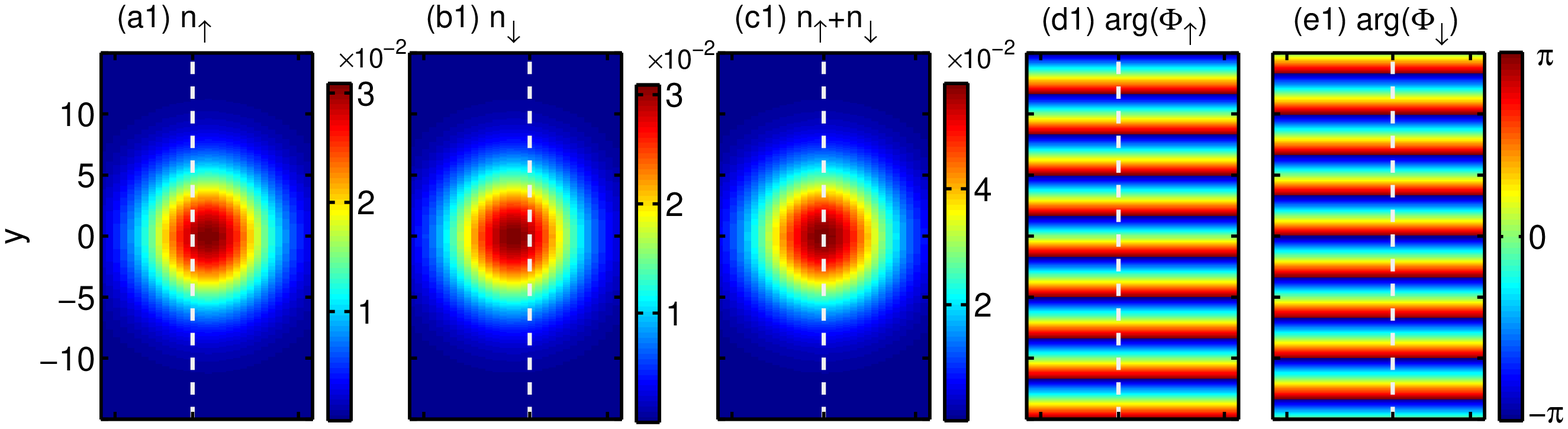}
\includegraphics[width=5.2in ]{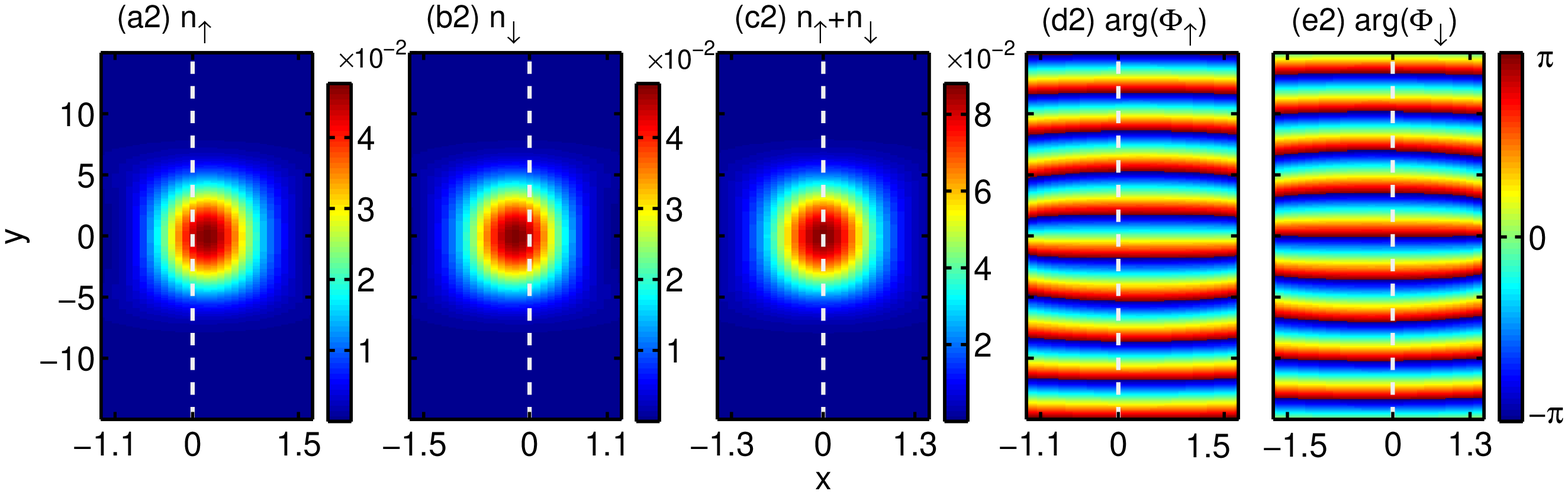}
\includegraphics[width=5.2in ]{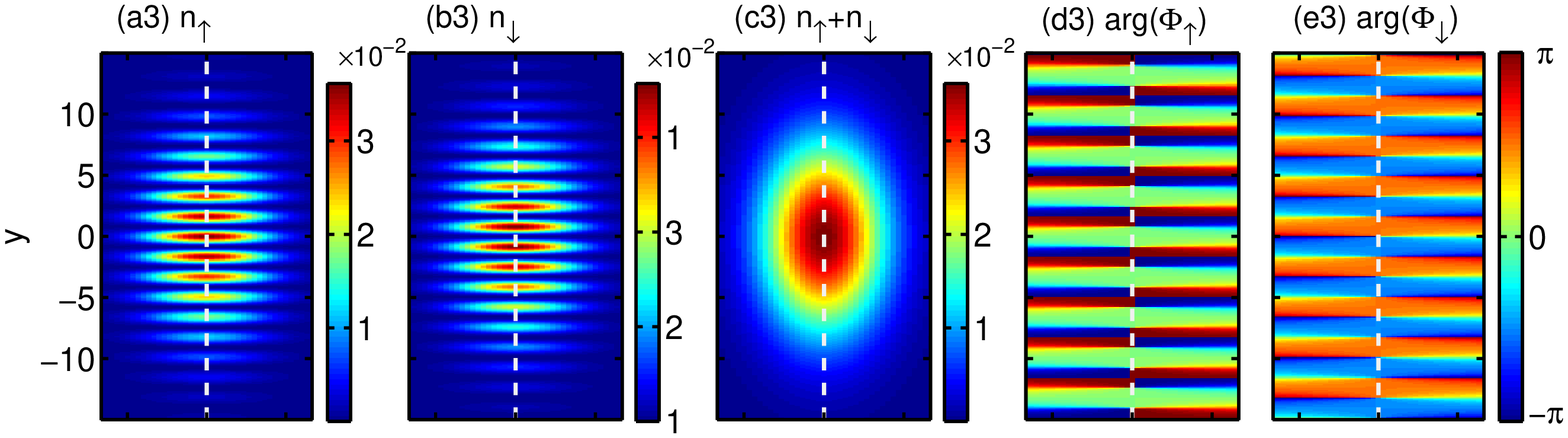}
\includegraphics[width=5.2in ]{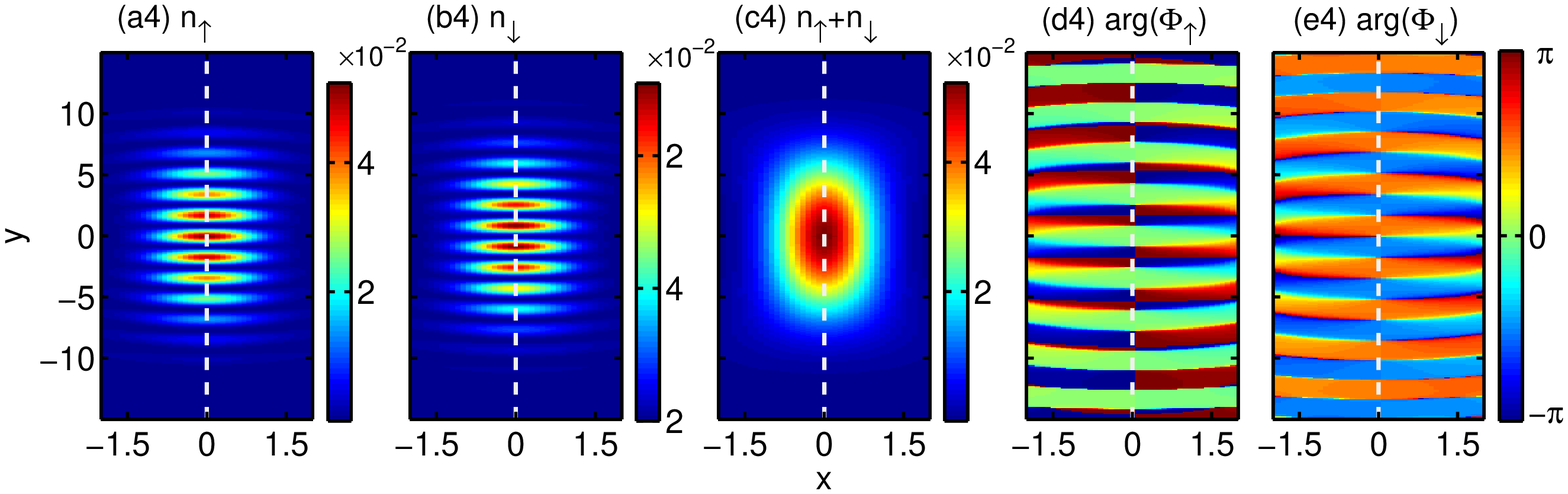}
\caption{(Color online) Profiles of the density $n_{\uparrow,\downarrow}=|\Phi_{\uparrow,\downarrow}|^2$
of spin $\uparrow$ ($\downarrow$) in (a)((b)),
the total density $n_\uparrow+n_\downarrow$ in (c), the phase of spin $\uparrow$($\downarrow$) in (d)((e)) for a plane
wave soliton (the first two panels) with $\gamma_{12}=6$ and
a stripe soliton (the last two panels) with $\gamma_{12}=10$. The solitons in the first and
third panels are obtained by the variational method, while the solitons in the second and forth panels
are calculated by the imaginary time evolution of the GP Eq.~(\ref{GP_time}). The dashed white line
labels the $x=0$ line. Here, $\gamma=8$, $\gamma_d/\gamma=0.67$, and $\alpha=2$.}
\label{wavefunction}
\end{figure*}

\section{Stationary bright solitons}
\label{StationarySoliton}
To shed light on the structure of a soliton, we start from the homogeneous noninteracting single particle scenario
and write its momentum space dispersion as
\begin{equation}
E({\bf k})=\frac{{\bf k}^2}{2}\pm\alpha k,
\end{equation}
with two branches labeled by the helicity $\pm$. Clearly, the ground state is degenerate with the energy
being $-\alpha^2/2$ when the momenta lie in the $k=|\alpha|$ ring. This is different from the case without spin-orbit coupling
where the ground state only occurs at $k=0$. In this single particle case, any superposition of the states
in the ring is also its ground state. Yet, this is not the case when the repulsive contact interaction is involved.
The ground state either possesses a single momentum (i.e. plane wave phase) when $\gamma_{12}/\gamma<1$
or two opposite momenta (i.e. stripe phase) when $\gamma_{12}/\gamma>1$~\cite{ZhaiHui2010PRL}. When the dipolar
interaction is turned on, one may expect that this effective long ranged attractive interaction along with contact
repulsive interaction could support two types of solitons: plane wave and stripe solitons.

\begin{figure}[t]
\includegraphics[width=3.4in]{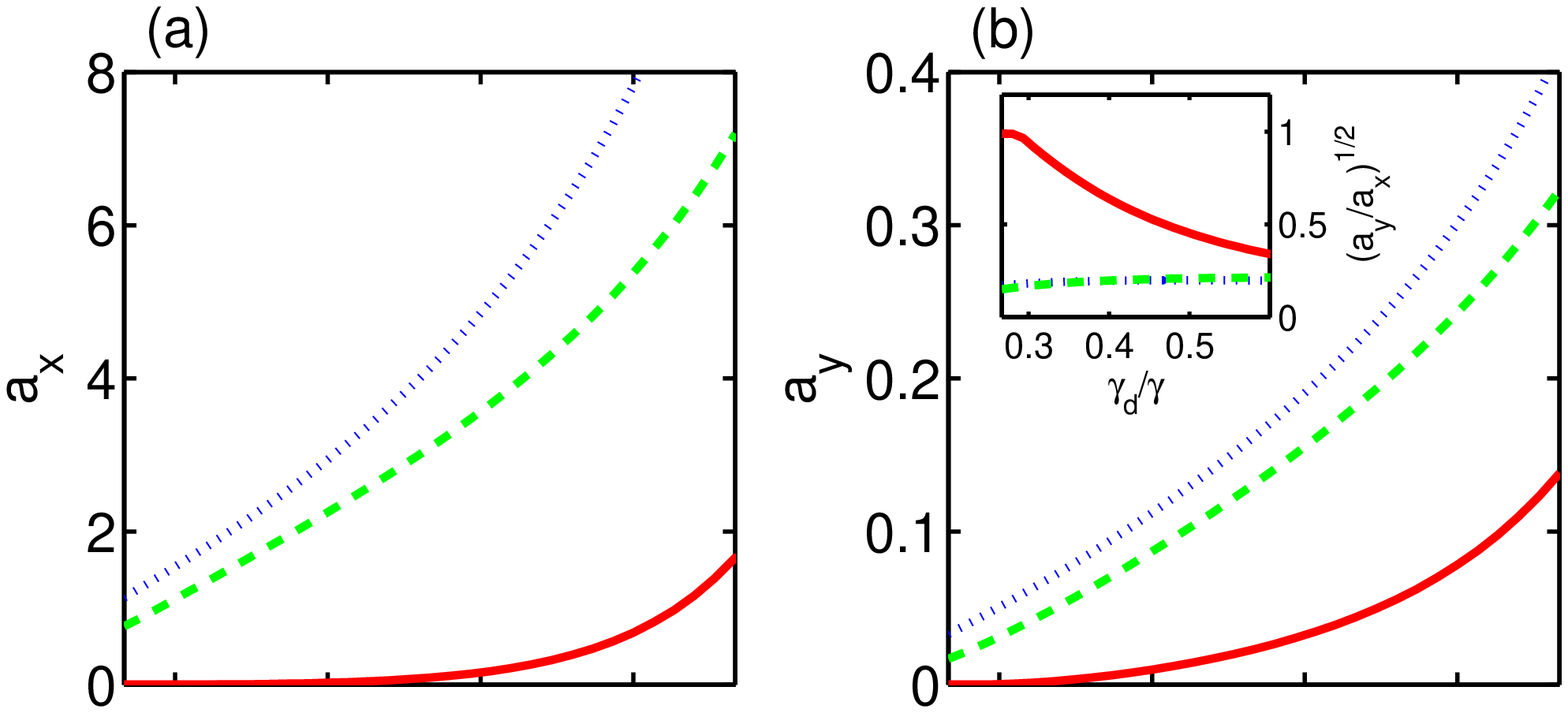}
\includegraphics[width=3.4in]{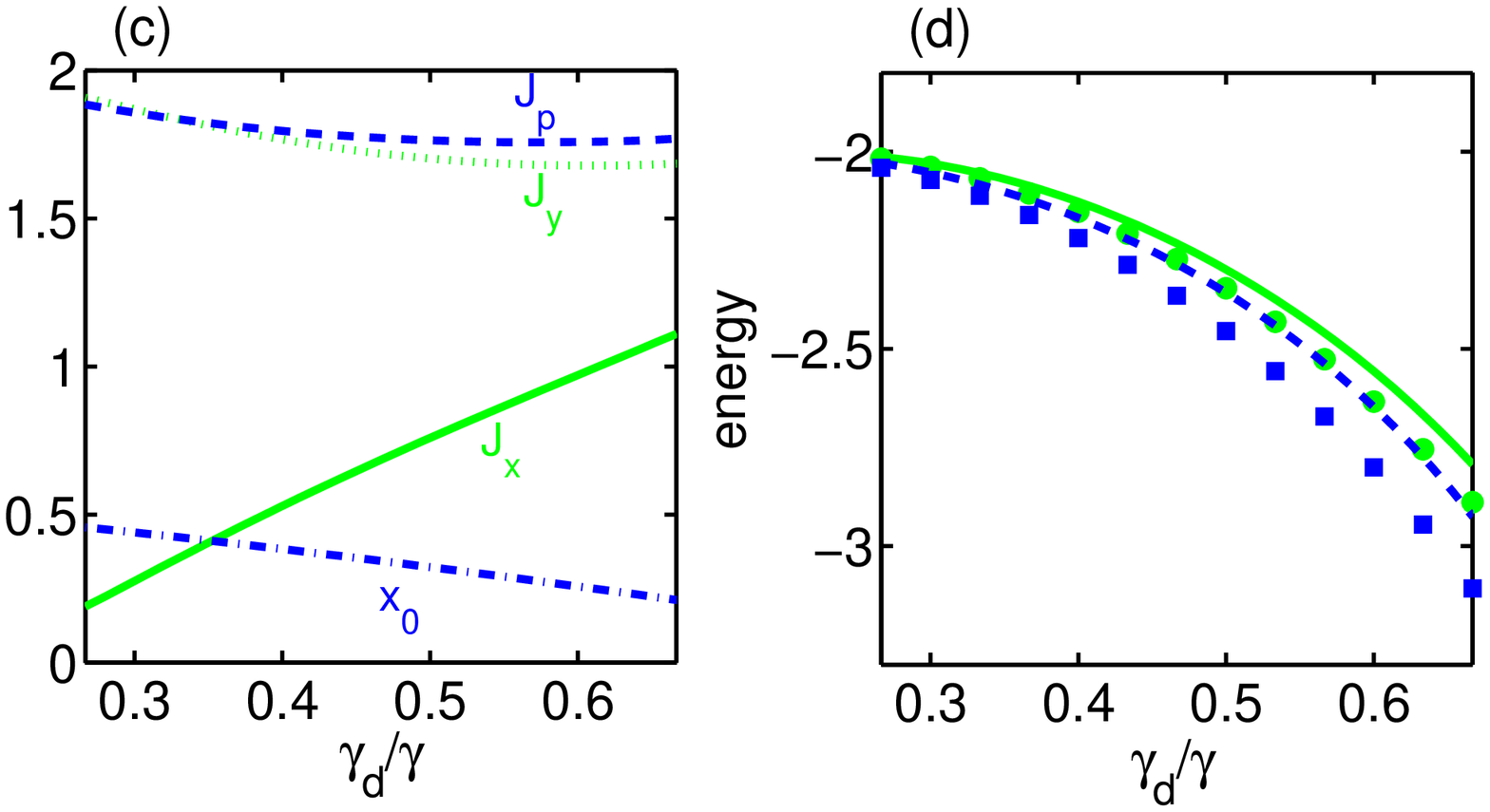}
\caption{(Color online) Plot of $a_x$ in (a) and $a_y$ in (b) as a function of $\gamma_d/\gamma$ for the
plane wave solitons (dotted blue line), stripe solitons (dashed green line), and traditional
solitons (solid red line) without spin-orbit coupling. The aspect ratio $\sqrt{a_y/a_x}$ of a soliton is
displayed in the inset of (b). Variational parameters with respect $\gamma_d/\gamma$ are
plotted in (c) associated with $x_0$ (dash-dot blue line) and $J_p$ (dashed blue line) for the plane
wave soliton, and $J_x$ (solid green) and $J_y$ (dotted green) for the stripe soliton.
In (d), the total energy of the variational ansatz wave function and the wave function numerically
obtained by the imaginary time evolution of the GPE for both plane wave and stripe solitons
is shown. The solid green line (stripe soliton) and dashed blue line (plane wave soliton) correspond to
the variational results, while the green circles and blue squares correspond to the GP results.
Here, $\alpha=2$, $\gamma=8$, $\gamma_{12}=6$ ($\gamma_{12}=10$) for the plane wave (stripe) soliton.
}
\label{functionAsgd}
\end{figure}

To examine whether a soliton can indeed exist in the spin-orbit-coupled dipolar BECs, we first consider
a plane wave soliton variational ansatz
\begin{eqnarray}
{\Phi_P}=
\begin{pmatrix}\Phi_0(x_0/2) \\ -\Phi_0(-x_0/2)\end{pmatrix}\text{exp}(-iJ_{p} y),
\label{ansatz_PW}
\end{eqnarray}
where
\begin{eqnarray}
\Phi_0(x_0)=\frac{(a_x a_y)^{1/4}}{\sqrt{2\pi}}e^{-\frac{1}{2}[a_x(x-x_0)^2+a_y y^2]}.
\end{eqnarray}
Here $J_p$ is the wave vector of the plane wave soliton,
$l_\nu=1/\sqrt{a_\nu}$ with $\nu=x,y$ is the size of the soliton, and $x_0$ is the separation distance between
two components. When $x_0=0$, this state is an eigenstate of $p_y\sigma_x$ multiplied by
a Gaussian profile $\Phi_0(0)$, and $J_p=\alpha$ yields the minimum energy.
In fact, $x_0$ is usually nonzero because
of a force acting on the BEC by spin-orbit coupling $\bf{F}=\alpha^2({\bf p}\times {\bf e}_z)
\sigma_z$~\cite{Yongping2012PRL,Shuwei2013JPB}, which is opposite along the $x$ direction
when $\bf p$ (here $J_p$) is along the $y$ direction.

In writing down the ansatz (\ref{ansatz_PW}), we have assumed that the wave vector $J_p$ is
in the $y$ direction. The prerequisite of this assumption is that the rotation
symmetry~\cite{Yongping2012PRL,Hu2012PRL} about the $z$ axis has been broken by the dipole-dipole interaction.
Indeed, without the dipole-dipole interaction, this state with $J_p$ along $y$ is not
special and other states with $J_p$ along other directions are degenerate with it.
For example, the state with $J_p$ along $y$ has the same energy as a state
with $J_p$ along $x$. Yet, with the specific dipole-dipole interaction arising from
the dipoles entirely oriented along $y$, the symmetry is broken and the ground state
should be elongated along $y$ ($a_x>a_y$) so as to provide an effective attractive
interaction because of the head-to-tail configuration of polarized dipoles. This elongated
configuration allows the existence of a 2D soliton~\cite{Malomed2008PRL} and also
requires the wave vector to be along $y$~\cite{note1}.

Although the wave vector $J_p$ of the ground state is along $y$, there are still two
options: negative and positive directions in terms of the time-reversal
symmetry $\mathcal{T}$ (i.e. $-i\sigma_y \mathcal{K}$ with the complex conjugate operator $\mathcal{K}$).
Specifically, the state $\Phi_{P2}=\mathcal{T}\Phi_P$ is degenerate with $\Phi_P$.
In the absence of interactions,
all superposition states of $\Phi_{P}$ and $\Phi_{P2}$,
\begin{equation}
\Phi_{PS}=|\cos\theta|\Phi_P+|\sin\theta| e^{i\varphi}\Phi_{P2},
\label{phisp}
\end{equation}
are degenerate. This degeneracy may be broken by the interaction so that the ground state
is either $\Phi_{P}$ or $\Phi_{P2}$, or a certain superposition state of them. But this degeneracy breaking should not happen
at $\gamma_{12}/\gamma=1$ since the interaction energy only depends on the total density which
is independent of $\theta$ and $\varphi$. This gives us an intuitive understanding
that $\gamma_{12}/\gamma=1$ may separate the plane wave soliton ($|\cos\theta|=$0 or 1) and
the stripe soliton ($|\cos\theta|=|\sin\theta|$), similar to the homogenous spin-orbit-coupled
BEC~\cite{ZhaiHui2010PRL} without dipole-dipole interactions. For the stripe soliton, we
note that $\varphi=0,\pi$ corresponds to
the ground state as the energy contributed by $\varphi$ is $-\gamma_{12}\sqrt{a_x a_y}e^{-(a_x x_0^2/2+2J_p^2/a_y)}\cos(2\varphi)/(16\pi)$
~\cite{ExplainVarphi}.

To evaluate the variational parameters $a_x$, $a_y$, $x_0$, $J_p$, and $\theta$, we minimize the energy $\mathcal{E}$
after substituting $\Phi_{PS}$ in Eq.~(\ref{phisp}) to Eq.~(\ref{energy_dim}). Indeed, the calculated
variational solutions reveal that there are two types of soliton
solutions: plane wave solitons when $\gamma_{12}/\gamma<1$ and stripe solitons when $\gamma_{12}/\gamma>1$.
We present the density and phase profiles of a typical plane wave soliton (we choose $\theta=\varphi=0$) in the first panel
of Fig.~\ref{wavefunction}, where the stripe structure of the phase of both two components
reveals the plane wave feature. The soliton is highly elongated along the $y$ direction and
the centers of two components are spatially separated along the $x$ direction because of nonzero $x_0$.
To confirm that this variational solution can qualitatively characterize the ground state of the
system, we numerically compute the ground state by an imaginary time evolution of the GP
Eq.~(\ref{GP_time}). This exact numerical solution also concludes that $\gamma_{12}/\gamma<1$
yields the plane wave soliton while $\gamma_{12}/\gamma>1$ the stripe soliton.
In the second panel of Fig.~\ref{wavefunction}, we also plot the corresponding density and phase profiles
of the GP obtained plane wave soliton. The variational ansatz is in qualitative agreement with it given
the separated centers and the plane wave varying phase that both states possess. Yet, the shape of
the soliton obtained by the imaginary time evolution deviates slightly from the Gaussian and the size
is also slightly smaller.

When $\theta=\pi/4$ and $\varphi=0$, $\Phi_{PS}$ is a stripe state with a density oscillation along
the $y$ direction for each component. And there is no stripe for the total density.
Along the $x$ direction, two components are not spatially separated, and the phase for
the spin $\uparrow$ reverses suddenly across $x=0$. Following these properties
by replacing $(\Phi_0(x_0/2)+\Phi_0(-x_0/2))/\sqrt{2}$ with $\cos(J_x x)\Phi_0(0)$ and
$(\Phi_0(x_0/2)-\Phi_0(-x_0/2))/\sqrt{2}$ with $\sin(J_x x)\Phi_0(0)$ in Eq.~(\ref{phisp}),
we obtain another better variational ansatz for the stripe soliton
\begin{equation}
\Phi_S=\Gamma \Phi_0(0),
\label{ansatz_stripe}
\end{equation}
where
\begin{eqnarray}
\Gamma=
\begin{pmatrix} \cos(J_{y} y)\cos(J_{x} x)-i\sin(J_{y} y)\sin(J_{x} x)
\\ \cos(J_{y} y)\sin(J_{x} x)+i\sin(J_{y} y)\cos(J_{x} x)\end{pmatrix},
\end{eqnarray}
with the variational parameters $J_x$ and $J_y$. The period of the stripe along
the $y$ direction is $\pi/J_y$. Interestingly, this stripe soliton
corresponds to four points $(\pm J_x,\pm J_y)$ in momentum space instead of
traditional two points~\cite{Yong2013PRA} when $J_x=0$, if we do not consider
the Gaussian profile $\Phi_0$.

We calculate the variational parameters of stripe solitons by performing the minimization of
the energy $\mathcal{E}$ in Eq.~(\ref{energy_dim}) where $\Phi$ is replaced with $\Phi_S$.
The density and phase profiles of a typical stripe soliton calculated by this method is
displayed in the third panel of Fig.~\ref{wavefunction}.
Evidently, the density of each component exhibits the stripe structure while
the total density does not. The phase of spin $\uparrow$ along the $y$ direction varies like
a plane wave, but reverses across $x=0$ due to the presence of $\sin(J_{x}x)$ in
the imaginary part. The phase of spin $\downarrow$ exhibits the phase rotation like vortices
around $x=0$ and $y=n\pi/J_y$ with integer $n$; around these points, the wave function $\Phi_{S\downarrow}$
is proportional to $(-1)^{n}(J_x x+i (J_y y-n\pi))$ and the corresponding density of spin
$\downarrow$ is extremely low. Moreover, in the last panel of Fig.~\ref{wavefunction}, we plot the density
and phase profiles of the corresponding stripe soliton obtained by the imaginary time evolution of the GPE;
comparing this figure with the third panel of Fig.~\ref{wavefunction} implies that the stripe
variational ansatz is qualitatively consistent with the GP results.

\begin{figure}[t]
\includegraphics[width=3.4in ]{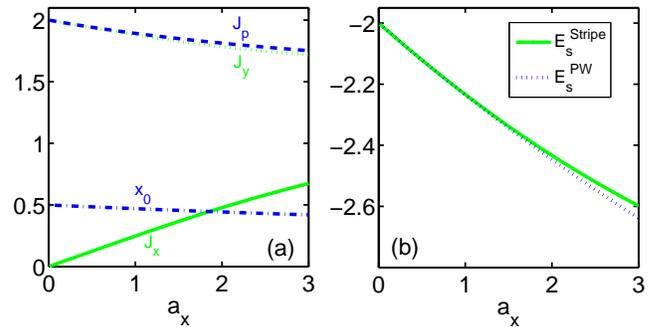}
\caption{(Color online) Plot of $x_0$ (dash-dot blue line) and $J_p$ (dashed blue line)
for the plane wave variational ansatz, and $J_x$ (solid green line) and
$J_y$ (dotted green line) for the stripe variational ansatz with respect to $a_x$
by the minimization of the energy $E_{s}^{\rm PW}$ and $E_{s}^{\rm Stripe}$ in (a). (b) shows
the minimum energy of $E_{s}^{\rm PW}$ (dotted blue line) and $E_{s}^{\rm Stripe}$ (solid green line)
as a function of $a_x$. Here,
$\alpha=2$.}
\label{singelEnergy}
\end{figure}

\begin{figure}[t]
\includegraphics[width=3.4in]{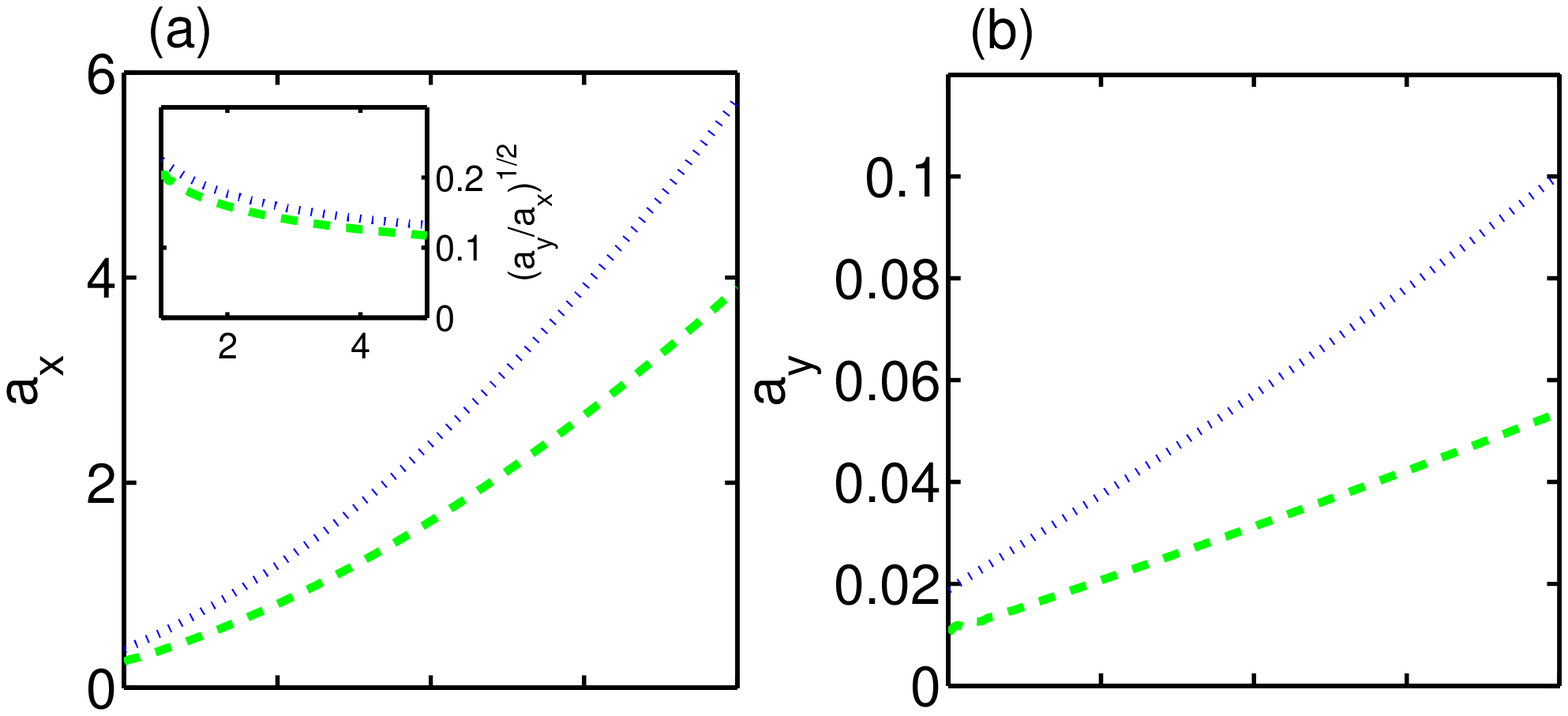}
\includegraphics[width=3.4in]{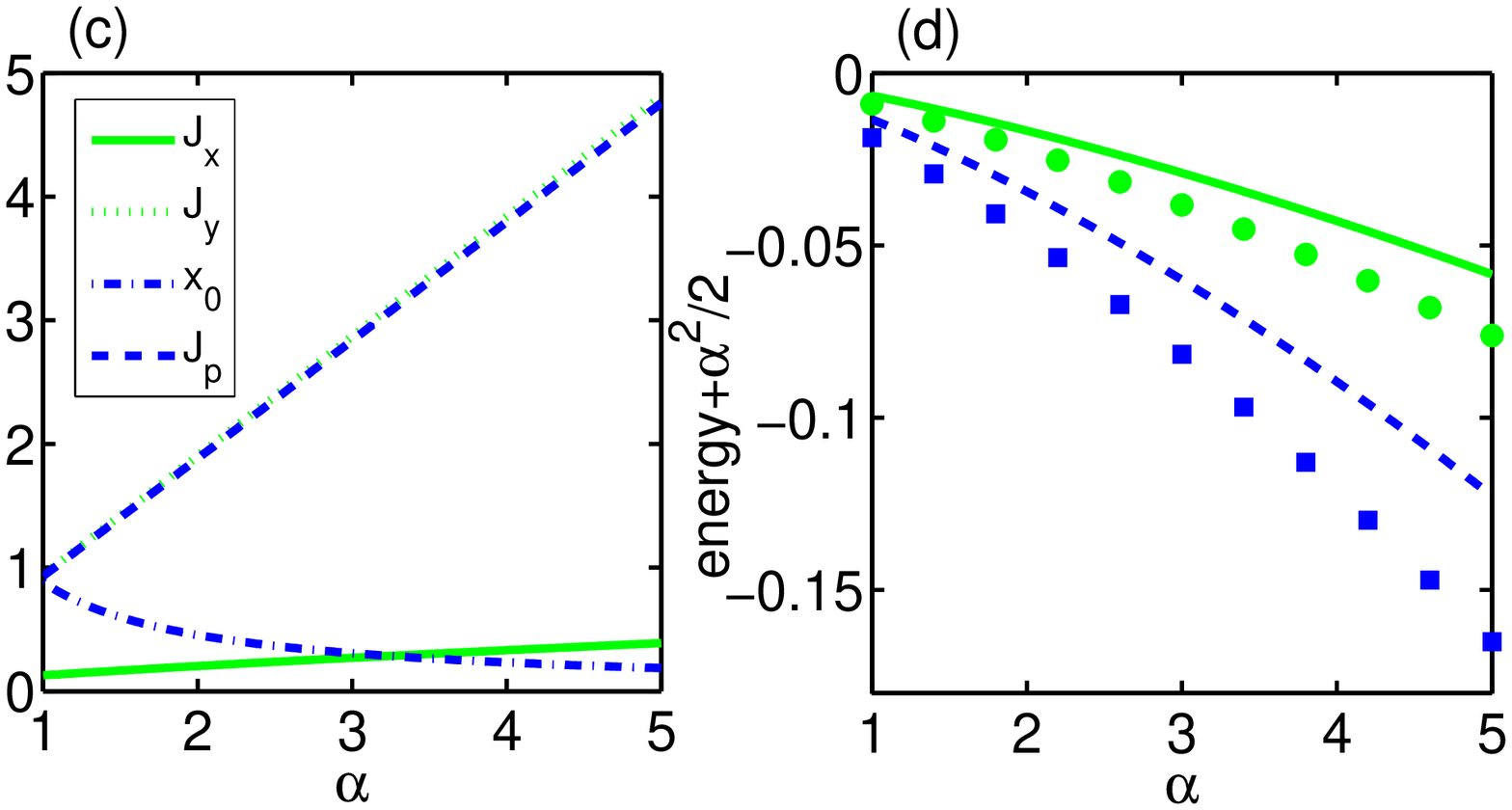}
\caption{(Color online) Plot of $a_x$ in (a) and $a_y$ in (b) as a function of the spin-orbit
coupling strength $\alpha$ for the plane wave soliton (dotted blue line) and stripe soliton
(dashed green line). The aspect ratio $\sqrt{a_y/a_x}$ of a soliton is plotted in the inset of (a).
(c) illustrates the change of $x_0$ (dash-dot blue line), $J_p$ (dashed blue line)
for the plane wave soliton and $J_x$ (solid green line) and $J_y$ (dotted green line)
for the stripe soliton with respect to $\alpha$. In (d), the total energy plus $\alpha^2/2$
is plotted as a function of $\alpha$. The solid green (for a stripe soliton) and dashed
blue (for a plane wave soliton) are calculated by the variational method, while
the green circles (for a stripe soliton) and blue squares (for a plane wave soliton)
are numerically obtained by the imaginary time evolution of the GPE. Here,
$\gamma=8$, $\gamma_d/\gamma=0.67$, and $\gamma_{12}=6$ ($\gamma_{12}=10$) for the plane wave (stripe) soliton.}
\label{functionAsalpha}
\end{figure}

To study the properties of a soliton with respect to dipole-dipole interactions $\gamma_d$, we evaluate
the variational parameters of both the plane wave and stripe solitons by the variational method
and plot them in Fig.~\ref{functionAsgd} as $\gamma_d/\gamma$ varies.
Clearly, with increasing $\gamma_d/\gamma$, $a_x$ and $a_y$ increase monotonously because of the
enhanced effective attractive interaction, indicating that the size $l_x$ and $l_y$ of the soliton
decrease monotonously.
We note that as $\gamma_d/\gamma$ increases further, the soliton can collapse so that both
$a_x$ and $a_y$ diverge. For the plane wave soliton, $a_x$ and $a_y$ are slightly larger
than the stripe soliton because of the smaller contact interaction of the former.
Moreover, compared with the soliton without spin-orbit coupling (red line in Fig.~\ref{functionAsgd}(a) and (b)),
$a_x$ and $a_y$ for both the plane wave and stripe solitons are much larger, implying that the size
of solitons can be reduced by spin-orbit coupling. Also, these solitons are highly anisotropic with the
much smaller aspect ratio $\sqrt{a_y/a_x}$ as shown in the inset of Fig.~\ref{functionAsgd}(b).
To elucidate the reason, we explicitly write that single particle energy of the plane wave
variational ansatz in Eq.(\ref{ansatz_PW}) which results from the presence of $x_0$ and $J_p$
\begin{equation}
E_s^{\rm PW}=\frac{1}{2}J_p^2-\alpha e^{-\frac{x_0^2 a_x}{4}} (J_p +
\frac{1}{2} a_x x_0).
\label{EsPW}
\end{equation}
The minimization of $E_s^{\rm PW}$ with respect to $x_0$ and $J_p$ for fixed $a_x$
yields
\begin{eqnarray}
x_0&=&\frac{-J_p+\sqrt{J_p^2+2a_x}}{a_x} \label{x0Eq} \\
J_p&=&\alpha e^{-x_0^2 a_x/4}. \label{JpEq}
\end{eqnarray}
For $a_x=0$, the energy is independent of $x_0$ and $J_p=\alpha$, while for $a_x\neq 0$, both $x_0$
and $J_p$ decrease slightly with increasing $a_x$ as shown in Fig.~\ref{singelEnergy}(a)
with the asymptotic $x_0=1/\alpha$ and $J_p=\alpha$ as $a_x$ goes zero. The energy
$E_{s}^{PW}$ is also a monotonously decreasing function of $a_x$. And this energy decline
combined with the reduced dipole-dipole interaction energy competes with the rise of
the kinetic energy (when $x_0=J_p=0$) and contact interaction energy, leading to
an increased $a_x$ and $a_y$ compared with the soliton without spin-orbit coupling.
This is also consistent with Fig.~\ref{functionAsgd}(c),
showing that with increasing the dipole-dipole interaction, $a_x$ increases and
both $x_0$ and $J_p$, therefore, decrease so as to lower $E_s^{\rm PW}$.
It is important to note that although $E_s^{\rm PW}$ is not a function of $a_y$,
other energy such as the kinetic energy (when $x_0=J_p=0$), the contact and
dipolar interaction energy depends on it.

For the stripe soliton, the single particle energy due to the presence of $J_x$ and $J_y$ is
\begin{equation}
E_s^{\rm Stripe}=\frac{1}{2}(J_{x}^2+J_{y}^2)-\alpha (J_{x}+
 J_{y} e^{-J_{x}^2/a_x}).
\end{equation}
Similar to the plane wave case, this energy is independent of $a_y$. For fixed $a_x$, the minimization
of this energy yields both $J_x$ and $J_y$ as a function of $a_x$ as shown in Fig.~\ref{singelEnergy}(a).
When $a_x$ moves towards zero, the solution approaches ($J_x=\alpha$, $J_y=0$) or ($J_x=0$, $J_y=\alpha$);
when it moves away from zero, there is only one solution where $J_y$ decreases from $\alpha$ while
$J_x$ increases from zero with the rise of $a_x$. Also, the energy $E_s^{\rm Strip}$ decreases
as $a_x$ increases. Analogous to the plane wave soliton, the total energy decrease resulted from spin-orbit coupling and
dipole-dipole interactions as $a_x$ and $a_y$ increase from the value without spin-orbit coupling
exceeds the energy gain of the kinetic (when $J_x=0$ and $J_y=0$) and contact interaction; this leads to the
increased $a_x$ and $a_y$ compared with the soliton without spin-orbit coupling.
This picture is also consistent with Fig.~\ref{functionAsgd}(c) where $J_x$ increases while $J_y$
decreases with respect to $\gamma_d/\gamma$.

To explicitly demonstrate the effect of the spin-orbit coupling on the properties of a soliton, we plot the variational parameters
as a function of the spin-orbit coupling strength $\alpha$ for both the plane wave and stripe solitons in
Fig.~\ref{functionAsalpha}. Consistent with the aforementioned feature that spin-orbit coupling can reduce the
size of the soliton, both Fig.~\ref{functionAsalpha}(a) and Fig.~\ref{functionAsalpha}(b) display a monotonous increasing behavior
of $a_x$ and $a_y$ as a function of $\alpha$. Also, the aspect ratio $\sqrt{a_y/a_x}$ is
decreased by spin-orbit coupling. Similar to Fig.~\ref{functionAsgd}(a) and Fig.~\ref{functionAsgd}(b), $a_x$ and $a_y$ for the plane
wave soliton are slightly larger than the stripe soliton in that the former has a smaller contact
interaction. For the plane wave soliton, $J_P$ (determined mainly by the spin-orbit coupling strength) increases
with respect to $\alpha$ while $x_0$ decreases; for the stripe soliton, both $J_x$ and $J_y$ increase.

In Fig.~\ref{functionAsgd}(d) and Fig.~\ref{functionAsalpha}(d), for both plane wave and stripe solitons,
we compare their energy obtained by the variational procedure with the one obtained by the imaginary time
evolution of the GPE. Both figures show that the energy calculated by the imaginary time evolution is lower
as expected. Yet, the difference between these two energy is not large (no more than 10\%), suggesting
that the variational ansatz can qualitatively characterize the solitons.
We note that in Fig.~\ref{functionAsalpha}(d), the energy is shifted by $\alpha^2/2$ in order to clearly present the
different results of the two methods, which could be smeared by the large value of $\alpha^2/2$.

\section{Moving bright solitons}
Generally, the wave function of a moving soliton with the velocity ${\bf v}$ can be
simply written as $\exp(i{\bf v}\cdot{\bf r})\Phi_s({\bf r}-{\bf v}t)$ where
$\Phi_s$ is the wave function of a stationary soliton. But this is only valid
for a system respecting Galilean transform invariance. In fact, Galilean
invariance is broken in a spin-orbit-coupled BEC~\cite{Qizhong2013}, and this violation dictates
that the shape of a soliton depends on its velocity strength~\cite{Yong2013PRA}.
Here, for a soliton in a spin-orbit-coupled dipolar BEC in 2D, we assume that a moving soliton can be written as
\begin{equation}
{\Phi_M}({\bf r},t)=\Phi_v({\bf r}-{\bf v} t, t)\exp({i{\bf v}\cdot{\bf r}-i\frac{1}{2} v^{2}t}),
\end{equation}
where $\Phi_v$ is a localized function. Plugging ${\Phi_M}({\bf r},t) $
into Eq. (\ref{GP_time}) yields
\begin{eqnarray}
i\frac{\partial\Phi_v({\bf r})}{\partial t}=&&\mathcal{H}_s({\bf v})\Phi_v({\bf r})+\gamma \mathcal{G}\Phi_v({\bf r}) \nonumber \\
&&+\gamma_d \int d{\bf k} e^{i{\bf k}\cdot{\bf r}}n({\bf k})U_d({\bf k}) \Phi_v({\bf r}),
\label{GP_time_v}
\end{eqnarray}
where $\mathcal{H}_s({\bf v})=\mathcal{H}_s+\alpha({\bf v}\times{\bm \sigma})\cdot{\bf e}_z$.
Compared to Eq.~(\ref{GP_time}), this dynamical equation has
an additional term $\alpha({\bf v}\times{\bm \sigma})\cdot{\bf e}_z$, acting as a Zeeman field;
this additional term implies the violation of Galilean invariance. This violation means that it is no
longer a trivial task to find a moving bright soliton for a BEC with spin-orbit coupling;
we need to perform an imaginary time evolution of the Eq.~(\ref{GP_time_v}),
but not Eq.~(\ref{GP_time}). Furthermore, such a 2D moving soliton should be different
for different velocity directions even if their amplitude is the same, in contrast
to a 1D soliton which can only move in one direction.

\begin{figure}[t]
\includegraphics[width=3.4in]{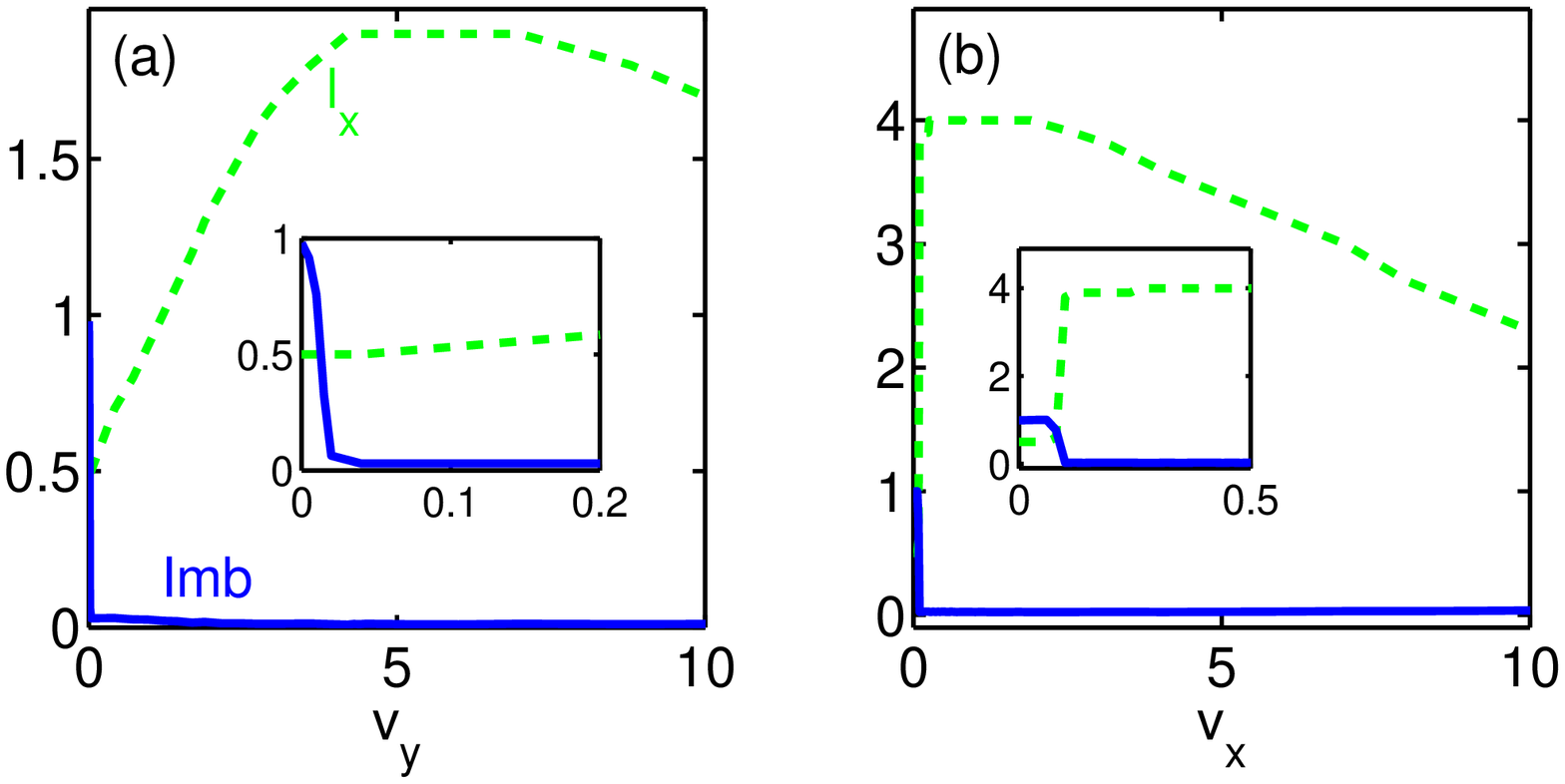}
\includegraphics[width=3.4in]{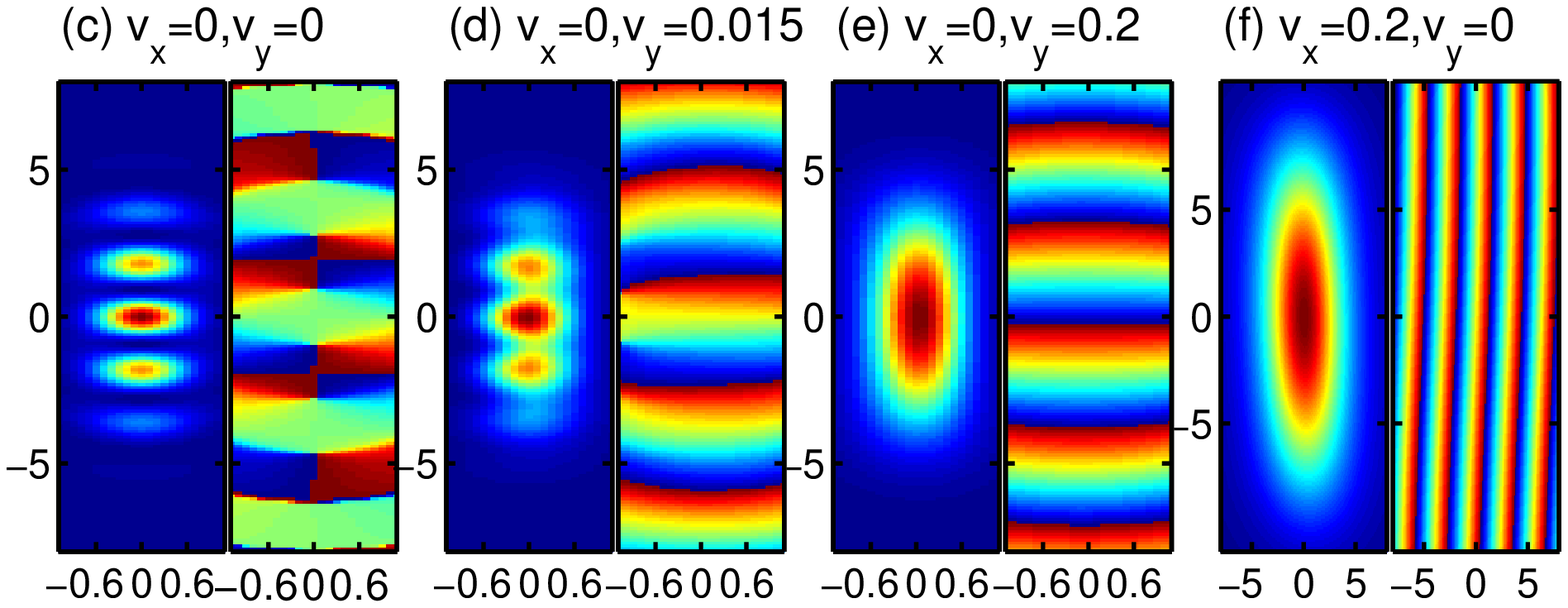}
\caption{(Color online) Imbalance $Imb$ and width $l_x$ of spin $\uparrow$ of
the solitons with respect to the velocity along the $y$ ($x$) direction in (a) ((b)).
The insets display the enlarged figure in a small velocity region.
Density and phase profiles of four typical moving solitons for spin $\uparrow$ corresponding to
the different velocities in (a) and (b) are plotted in (c-f) where the horizontal and vertical coordinates
are $x$ and $y$ respectively.
Here, $\alpha=2$, $\gamma=8$, $\gamma_{12}=10$, and $\gamma_d/\gamma=1$,
corresponding to a stripe soliton when stationary.
}
\label{MovingSoliton}
\end{figure}

To examine how the shape of a soliton changes with respect to the velocities along $x$ and $y$
directions, we plot the imbalance $Imb$ and the width $l_x$ of a soliton of spin $\uparrow$
as a function of the velocities $v_x$ and $v_y$ in Fig.~\ref{MovingSoliton}.
Here, the imbalance for spin $\uparrow$ is defined as
\begin{equation}
Imb=\frac{|\Phi_{\uparrow}(0)|^2-
|\Phi_{\uparrow}(\pi/(2\alpha))|^2}{|\Phi_{\uparrow}(0)|^2+|\Phi_{\uparrow}(\pi/(2\alpha))|^2},
\end{equation}
which characterizes a stripe soliton (as shown in Fig.~\ref{MovingSoliton}(c) and Fig.~\ref{MovingSoliton}(d))
when it approaches one and a plane wave soliton (as shown in Fig.~\ref{MovingSoliton}(e) and
Fig.~\ref{MovingSoliton}(f)) when it approaches zero.
Fig.~\ref{MovingSoliton}(a) and Fig.~\ref{MovingSoliton}(b) demonstrate that
$Imb$ suffers a sharp decline from one to near zero as $v_x$ and $v_y$ increase, indicating
that a moving soliton tends to be a plane wave state. The reason is the broken rotation symmetry
of the single particle Hamiltonian by the velocity induced Zeeman field, giving rise to a ground state of
the single particle system lying at one momentum point located along the $x$ ($y$) direction when the
velocity is along that direction. This also explains why the phase of a moving plane wave soliton with
the velocity along the $x$ ($y$) direction varies along that direction.

Furthermore, Fig.~\ref{MovingSoliton}(a) demonstrates that the
width of the soliton gradually grows when the velocity along the $y$ direction is enlarged,
To explain the growth,
we consider the plane wave ansatz in Eq.~(\ref{ansatz_PW}) which yields an additional
term $-\alpha v_y e^{-a_x x_0^2/4}$ for the single particle energy when a soliton moves;
this energy decrease enlarges exponentially with the decline of $a_x$ (i.e. increase
of the width), leading to an expanded soliton with the rise of the velocity.
However, this is not a monotonous behavior and the soliton begins shrinking when the velocity
goes larger, due to the enlarged $J_p$ by the velocity induced Zeeman field, similar to
increasing spin-orbit coupling. On the other hand,
when the velocity is along the $x$ direction, the width of the soliton gains a sudden rise as the velocity
varies, as shown Fig.~\ref{MovingSoliton}(b). This corresponds to a change from a stripe soliton
with the wave vector along the $y$ direction to a plane wave soliton with the wave vector
along the $x$ direction. For the stationary solitons, the soliton with the wave vector mainly
along the $y$ direction has lower energy than the one with the wave vector mainly along the
$x$ direction as the dipoles are completely oriented along $y$. But
the Zeeman field induced by the presence of a velocity along the $x$ direction gives rise to the
single particle ground state that possesses the wave vector along $x$. The two states with the wave
vector along these two directions compete and change from the former to the latter (i.e.
first order phase transition). For the decrease of the width when $v_x$
goes even larger, the reason is the same as the case for $v_y$. When a stationary soliton is
plane wave, the moving behavior is similar except that the moving soliton is always the plane
wave soliton.

\section{Conclusion}
We have studied the bright solitons as the ground states in a spin-orbit-coupled dipolar BEC in 2D
with dipoles completely polarized along one direction in the 2D plane. It is important to note that the solitons
are the ground states in 2D, but they are the metastable states in quasi-2D where the true
ground state would collapse and there is an energy barrier between the soliton state and this
ground state. Two types of solitons have been found: a plane wave soliton and a stripe soliton.
The former has the plane wave phase variation and its two components are slightly spatially
separated; while for the latter, the density of each component is spatially oscillating and
the variational ansatz suggests that four points in momentum space are involved.
Both plane wave and stripe solitons are highly anisotropic and their size is decreased by
spin-orbit coupling. These solitons cannot exist as the ground states in a 2D system with pure attractive contact interactions and
spin-orbit coupling. Moreover, the shape of these solitons changes with their velocities due to the absence of
Galilean invariance, and this change is anisotropic.

The 2D bright soliton, albeit mainly plane wave soliton, can also
exist when equal Rashba and Dresselhaus spin-orbit coupling is considered. In experiments, this
type of spin-orbit coupling has been engineered by coupling two hyperfine states of atoms through
two counterpropagating Raman laser beams~\cite{Lin2011Nature,Jing2012PRL,Zwierlen2012PRL,PanJian2012PRL,Peter2013,Spilman2013PRL}
and such setup could be employed to realize this spin-orbit coupling in Dysprosium~\cite{Cui2013PRA} with large
dipole-dipole interactions. Also, the large magnetic moment in Dysprosium atoms may permit the realization of Rashba spin-orbit
coupling~\cite{Spielman2011PRA}.

\begin{acknowledgments}
We would like to thank L. Jiang, K. Sun, C. Qu, Z. Zheng, and L. D. Carr for helpful discussions.
Y. Xu and C. Zhang are supported by ARO(W911NF-12-1-0334) and AFOSR (FA9550-13-1-0045).
Y. Zhang is supported by Okinawa Institute of Science and Technology Graduate University.
We also thank Texas Advanced Computing Center as parts of our numerical
calculations were performed there.
\end{acknowledgments}


\begin{thebibliography}{99}
\bibitem{note0}{Both the kinetic energy and the contact attractive interaction energy in 2D is proportial to $1/L^2$ with the size $L$ of a
state; to lower energy, the state collapses (i.e. $L \rightarrow 0$) when the interaction energy is larger than the kinetic one, and the
state expands (i.e. $L \rightarrow 0$) otherwise. }
\bibitem{note1}{This elongated configuration along with the wave vector in the $y$ direction has the lower
single particle energy contributed by the spin-orbit coupling than the case with the wave vector in
the $x$ direction.}
\end{thebibliography}
\end{document}